\documentclass[hidelinks]{statsoc}

\usepackage{natbib,color,hyperref,bm,amsmath,amsthm,amssymb,graphicx,float,booktabs,multirow,threeparttable,textcomp,times,accents,bbm,ragged2e,framed,algorithm2e,setspace,easyReview}
\usepackage[justification=raggedright,singlelinecheck=false]{subcaption}
\usepackage{tikz}
\usetikzlibrary{arrows.meta,positioning,shapes.geometric}
\usepackage{easyReview}
\usepackage{overpic}

\theoremstyle{plain}
\newtheorem{theorem}{Theorem}

\newtheorem{assumption}{Assumption}
\newtheorem{proposition}{Proposition}
\newtheorem{example}{Example}

\makeatletter
\def\IF{{\rm  IF}}

\definecolor{darkblue}{rgb}{0.0,0.0,0.6}

\def\t{{ \mathrm{\scriptscriptstyle T} }}
\def\pr{ f}

\def\expit{\text{\rm expit}}

\makeatother

\makeatletter
\newcommand*{\ind}{%
	\mathbin{%
		\mathpalette{\@ind}{}%
	}%
}
\newcommand*{\nind}{%
	\mathbin{
		\mathpalette{\@ind}{\not}
	}%
}
\newcommand*{\@ind}[2]{%
	\sbox0{$#1\perp\m@th$}
	\sbox2{$#1=$}
	\sbox4{$#1\vcenter{}$}
	\rlap{\copy0}
	\dimen@=\dimexpr\ht2-\ht4-.2pt\relax
	\kern\dimen@
	{#2}%
	\kern\dimen@
	\copy0 
} 
\makeatother

\usepackage{geometry}
\usepackage{xr}
\graphicspath{{./Rscripts/}}

\title{\bf Identifying the desert decision rule to assess and achieve fairness}
\author{Ping Zhang, Naiwen Ying and Wang Miao\\
Department of Probability and Statistics, Peking University, 5 Summer Palace Road, Haidian District, Beijing 100871, China}

\begin{document}

\begin{abstract}
We study fairness in decision-making when the data may encode systematic bias.
Existing approaches typically impose fairness constraints while predicting the observed decision, which may itself be unfair.
We propose a novel framework for characterising and addressing fairness issues by introducing the notion of desert decision, a latent variable representing the decision an individual rightfully deserves based on their actions, efforts, or abilities.
This formulation shifts the prediction target from the potentially biased observed decision to the desert decision.
We advocate achieving fair decision-making by predicting the desert decision and assessing unfairness by the discrepancy between desert and observed decisions.
We establish nonparametric identification results under causally interpretable assumptions on the fairness of the desert decision and the unfairness mechanism of the observed decision.
For estimation, we develop a sieve maximum likelihood estimator for the desert decision rule and an influence-function-based estimator for the degree of unfairness.
Sensitivity analysis procedures are further proposed to assess robustness to violations of identifying assumptions.
Our framework connects fairness with measurement error models, aligning predictive accuracy with fairness relative to an appropriate target, and providing a structural approach to modelling the unfairness mechanism.
\end{abstract}

\pagestyle{plain}
\setstretch{1.9}

\section{Introduction}

Statistical and machine learning methods have achieved substantial success in recent years and are now widely deployed in domains including hiring \citep{pessach2020employees}, lending \citep{thomas2017credit}, medicine \citep{shehab2022machine}, education \citep{luan2021review}, and criminal justice \citep{brennan2009evaluating}. Decisions that once relied primarily on human judgement are increasingly supported by predictive models. Alongside these advances, concerns regarding fairness have become more prominent \citep[e.g.,][]{nature2016,kleinberg2018discrimination,vanderford2022}. Unfairness often arises as systematic differences in treatment or outcomes across demographic groups. A widely cited example is a risk assessment tool used in the U.S. criminal justice system, which has been reported to assign higher predicted recidivism risk to Black defendants than to White defendants \citep{angwin2022machine}. Such disparities can shape individuals’ life chances and entrench existing social inequities. Ensuring fairness should therefore be treated not as a peripheral adjustment, but as a central requirement for statistical and machine learning methods that are deployed in socially consequential settings.

Despite widespread concerns regarding algorithmic fairness, most statistical and machine learning methods are, in themselves, neutral and are not designed to discriminate intentionally against particular individuals or groups. Rather, unfairness commonly originates in the training data: recorded decisions may reflect historical biases, including discrimination against minority groups. For example, in the labour market, Black applicants may face greater barriers to employment than White applicants even when they have comparable skills or qualifications \citep{bertrand2004emily}. When a model is trained to predict the observed decision, it may learn and propagate the bias, thereby producing unfair or unwarranted predictions. Consequently, in settings where unfairness is likely to be present, a fair decision rule should not merely replicate the observed decision as a predictive target. Among philosophical accounts of fairness, one influential view holds that fairness is achieved when each individual receives the decision they rightfully deserve, given their actions, efforts, or abilities; see the literature on desert \citep{feinberg1970justice, sadurski1985giving, miller2001principles}. Under this perspective, the appropriate target is the \textit{desert decision} rather than the observed decision, with the latter acting as a potentially biased proxy for the former.

The extensive literature on algorithmic fairness has produced a wide range of criteria and methods for improving fairness in decision-making, suited to different application settings; see \citet{mehrabi2021survey,mitchell2021algorithmic,barocas2023fairness,he2025fairness} for comprehensive reviews.
Despite their differences, most approaches share a common objective: to predict the observed decision while enforcing a pre-specified fairness criterion.
Enforcing fairness constraints can reduce disparities in predicted decisions, but the target remains the observed decision, which may itself be unfair.
Consequently, efforts to improve fairness often entail a reduction in predictive accuracy, yielding the well-recognised utility--fairness trade-off \citep{kleinberg2016inherent,zhao2022inherent}. Moreover, without an explicit characterisation of the mechanism generating unfairness, imposing fairness constraints while predicting the observed decision may produce a form of “balance” that lacks a clear causal interpretation and substantive justification in practice. A more principled alternative is to shift the target from the observed decision to the desert decision, which is intended to be fair. Under this shift, fairness is preserved by construction while predictive accuracy is pursued with respect to an appropriate target, thereby reducing the need to trade utility against fairness.

In this paper, we propose a novel framework for assessing and achieving fairness centred on the desert decision. 
The paper has three main contributions:
\begin{itemize}

\item  \textit{Conceptual contribution.} 
We introduce the desert decision as the prediction target and define fairness at the level of the desert decision. 
This perspective shifts the central question from ``how to predict the observed decision while enforcing a fairness constraint'' to ``how to recover and predict the desert decision'', so that fairness and predictive accuracy are aligned by construction.
In this framework,  unfairness is quantified by the discrepancy between the desert and observed decisions rather than by constraints imposed on the observed decision or model output.

\item  \textit{Identification and estimation methods.} Because the desert decision is unobserved, identification is challenging.
We establish identification of the desert decision rule and the degree of unfairness,
under transparent and causally interpretable assumptions on (i) the fairness property of the desert decision, (ii) the mechanism through which unfairness distorts the observed decision, and (iii) an auxiliary variable that is informative about the desert decision but does not affect the unfairness mechanism. 
We then develop a sieve maximum likelihood estimator for the desert decision rule and an influence-function-based estimator for the degree of unfairness, together with sensitivity analysis procedures for assumption violations.

\item \textit{Unfairness mechanism.} We explicitly characterise and estimate the mechanism of unfairness, providing a structural interpretation of discrimination and preferential treatment. 
In contrast, although the mechanism of unfairness is sometimes discussed in specific examples \citep{barocas2023fairness}, much of the algorithmic fairness literature does not formally define, identify, or estimate an explicit unfairness-generating mechanism.
\end{itemize}
Besides, this work    also contributes new identification and estimation methods to the  measurement error literature.

The remainder of the paper is organised as follows. In Section~\ref{sec:framework}, we review existing fairness criteria and methods, introduce the notion of the desert decision, and define parameters for assessing the degree of unfairness and for achieving fair decision-making. In Section~\ref{sec:iden}, we describe and illustrate assumptions on the mechanism of unfairness and establish identification results. Estimation methods for the parameters of interest are developed in Section~\ref{sec:est}. In Section~\ref{sec:sen}, we discuss extensions that accommodate violations of the identifying assumptions and propose sensitivity analysis procedures. For illustration, we conduct numerical simulations in Section~\ref{sec:sim} and apply the proposed methods to a real-data example \citep{bertrand2004emily} in Section~\ref{sec:app}. Section~\ref{sec:dis} concludes and outlines directions for future work.

\setstretch{1.8}

\section{Fairness and the desert decision framework} \label{sec:framework}
\subsection{Existing fairness criteria and methods}
Throughout the paper, let $Y$ denote the observed decision, with $Y=1$ indicating the favourable decision (e.g., hiring, loan approval) and $Y=0$ the unfavourable decision.
Let $S$ denote the sensitive attribute, with $S=1$ for the advantaged group and $S=0$ for the disadvantaged group.
Here, the advantaged group refers to individuals who are historically privileged (e.g., men, ethnic majority groups), whereas the disadvantaged group refers to those who typically face discrimination (e.g., women, ethnic minority groups).
Let $V$ denote a vector of nonsensitive covariates that are relevant for decision-making.
We use $\pr$ to denote a generic probability mass or density function.
The observed decision $Y$ may be subject to unfairness with respect to the sensitive attribute $S$.
For example, $Y$ may depend on $S$ (i.e., $Y \nind S$), or it may remain associated with $S$ even after conditioning on explanatory characteristics (i.e., $Y \nind S \mid V$).

We briefly review representative fairness criteria and methods in the algorithmic fairness literature.
Let $d(S,V)$ denote a decision rule or score taking values in $[0,1]$.
The predicted decision induced by $d(S,V)$ is $\hat Y = I\{d(S,V) > t\}$, where $I(\cdot)$ denotes the indicator function and $t$ is a given threshold.
Table~\ref{tbl:fair} summarises widely used fairness criteria expressed as statistical independence conditions.
Statistical parity and conditional statistical parity require independence between the predicted decision and the sensitive attribute \citep{dwork2012fairness,kamiran2013explainable}.
Equalised odds and equal opportunity require the predicted decision to be independent of the sensitive attribute conditional on the observed decision \citep{hardt2016equality}.
Calibration and predictive parity require independence between the observed decision and the sensitive attribute given the score or predicted decision \citep{chouldechova2017fair}.
In addition to criteria based on independence, causal notions of fairness have been proposed, including path-specific fairness \citep{zhang2017causal,nabi2018fair}, counterfactual fairness \citep{kusner2017counterfactual}, and principal fairness \citep{imai2023principal}, each corresponding to the absence of certain causal effects.
These criteria provide useful perspectives on different aspects of fairness, tailored to different application settings.

\begin{table}[H]
\centering
\caption{Definitions of fairness criteria}\label{tbl:fair}
\begin{tabular}{ll}
\toprule
Criterion & Definition \\
\midrule
Statistical Parity \citep{dwork2012fairness}
& $\hat Y \ind S$ \\
Conditional Statistical Parity \citep{kamiran2013explainable}
& $\hat Y \ind S \mid V$ \\
Equalised Odds \citep{hardt2016equality}
& $\hat Y \ind S \mid Y$ \\
Equal Opportunity \citep{hardt2016equality}
& $\hat Y \ind S\mid Y=1$ \\
Calibration \citep{chouldechova2017fair}
& $Y\ind S\mid d(S,V)$ \\
Predictive Parity \citep{chouldechova2017fair}
& $Y\ind S\mid \hat Y=1$ \\
\bottomrule
\end{tabular}
\end{table}

Representative fairness-enhancing methods include: adjusting the training data to mitigate bias \citep{kamiran2012data,calmon2017optimized,chen2024learning}; incorporating fairness requirements into model training via regularisations or constraints \citep{donini2018empirical,zafar2019fairness}; and modifying the outputs of trained models by group-specific decision thresholds for sensitive attributes \citep{hardt2016equality,fan2023neyman}.
Despite differences in implementation, these approaches typically share the objective of predicting the observed decision $Y$ while enforcing fairness.
This objective can be formulated as the following constrained statistical learning problem \citep{nabi2024statistical}:
\[\underset{d\in\mathcal D: C(d)=0}{ \arg\min}  \ R(d),\]
where $R(d)$ is a risk function for predicting $Y$, $C(d)=0$ is the constraint induced by a chosen fairness criterion, and $\mathcal D$ denotes the set of measurable functions of $(S,V)$ taking values in $[0,1]$.
For example, choosing the constraint $C(d) = E\{d(1,V)-d(0,V)\}$, which enforces a zero average causal effect of $S$ on $Y$, \citet{nabi2024statistical} derived the best predictors of $Y$ under various risk functions.
See the supplementary material for further details.
Decision rules obtained from such constrained learning problems can reduce disparities in decisions.
However, when unfairness is present, predicting the observed decision $Y$ may be an inappropriate target; the goal should instead be to recover the desert decision, rather than the potentially unfair decision $Y$ itself.

\subsection{Desert decision and parameters of interest}

We introduce the desert decision $Y^*$, a latent variable representing the deserved outcome that accords with an individual's actions, efforts, or abilities, drawing on the notion of ``desert'' in philosophy \citep[e.g.,][]{feinberg1970justice, sadurski1985giving, miller2001principles}.
In the algorithmic fairness literature, the concept of desert has been discussed as a general fairness principle \citep{barocas2023fairness}; in this paper, we define the desert decision as an individual-level variable.
Unlike the observed decision $Y$, which may be distorted by unfairness, the desert decision $Y^*$ corresponds to the decision that would have been made in the absence of unfairness, that is, the appropriate decision in an idealised setting in which individuals receive what they rightfully deserve.
However, the desert decision is not realised in the observed world, or at least cannot be verified as such.
To further illustrate the concept of the desert decision, we provide an example based on bank loan approval using credit scoring \citep{hand1997statistical,thomas2017credit}.

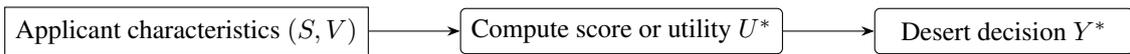
\begin{figure}[H]
\centering
\begin{tikzpicture}[
  node distance=8mm and 12mm,
  >=Stealth,
  every node/.style={font=\sffamily},
  process/.style={rectangle, draw, rounded corners=2pt, minimum width=34mm, minimum height=6mm, align=center, font=\small},
  data/.style={rectangle, draw, align=center, font=\small},
  group/.style={draw, rounded corners=4pt, inner sep=4pt, thick}
]

\node[data] (applicant) {Applicant characteristics $(S,V)$};
\node[process, right=of applicant] (score) {Compute score or utility $U^*$};
\node[process, right=of score] (desert) {Desert decision $Y^*$};
\draw[->] (applicant) -- (score);
\draw[->] (score) -- (desert);
\end{tikzpicture}
\caption{Desert decision in a standard loan approval procedure.}\label{fig:credit}
\end{figure}

\begin{example}\label{ex:desert}
Figure~\ref{fig:credit} illustrates a loan approval procedure whereby a bank collects applicant characteristics, computes a score based on a predefined rule, and approves the loan when the score exceeds a threshold.
Formally,
the desert decision is characterised by the following generative model: 
\begin{eqnarray*}
U^* = g(V,\varepsilon^*),\quad Y^* = I(U^* > 0),
\end{eqnarray*}
where the function $g$ stands for a pre-specified rule,
$\varepsilon^*$ is an independent error term,
and $U^*$ represents the score minus the threshold of loan approval.
This procedure can also be interpreted through discrete choice theory \citep{mcfadden2001economic, train2009discrete}, where $U^*$ represents the utility of approving the loan for an applicant with nonsensitive characteristics $V$, such as annual income, debt-to-income ratio, and the number of past overdue payments.
Importantly, the utility for the desert decision only depends  on nonsensitive characteristics that are associated with default risk;
and legislation explicitly prevents the use of sensitive attribute $S$ such as gender or race \citep{hand1997statistical}.
\end{example}

If the decision rule defining the desert decision were implemented faithfully in practice, the decision recorded in the data would be independent of $S$ given $V$. However, in practice the loan approval process may be contaminated by unfairness: 
the realised decision may be influenced by the sensitive attribute $S$, and hence may deviate from the desert decision. The desert decision is intended to be both fair and substantively justified; we encode its fairness property through the following assumption.

\begin{assumption}\label{ass:fair}
$Y^* \ind S \mid V$.
\end{assumption}

Assumption~\ref{ass:fair} requires the desert decision $Y^*$ to be conditionally independent of $S$ given $V$. Its plausibility is supported by a legal characterisation of employment discrimination \citep{7thCircuit1996}: ``The central question in any employment-discrimination case is whether the employer would have taken the same action had the employee been of a different race (age, sex, religion, national origin, etc.) and everything else had been the same''. 
In the bank loan Example~\ref{ex:desert}, the desert decision does not depend on the sensitive attribute, so Assumption~\ref{ass:fair} holds. 
We emphasise that this assumption is a basic requirement but not sufficient to characterise $Y^*$; for example, a coin-flip decision also satisfies the conditional independence yet cannot represent what individuals rightfully deserve.

To achieve fair decision-making, we advocate basing decisions on predictions of the desert decision $Y^*$. Specifically, we consider the decision rule
\[\tau(V) = \pr(Y^*=1\mid V),\]
the conditional probability of the favourable desert decision given the predictive covariates. 
The desert decision rule $\tau(V)$ has the following property.

\begin{proposition}\label{prop:pro}
Under Assumption~\ref{ass:fair},

(i) $\tau(V)$ is the best predictor of the desert decision under the cross-entropy risk:
\begin{align*}
\tau(V)  = \underset{d \in \mathcal D}{\arg\min} \ E[-Y^*\log d(S,V) - (1-Y^*)\log\{1-d(S,V)\}];
\end{align*}

(ii) $\tau(V)$ satisfies calibration for predicting the desert decision:
$Y^* \ind S \mid \tau(V).$
\end{proposition}

Although Proposition~\ref{prop:pro} is immediate, 
its implications are substantive. 
First, $\tau(V)$ attains the minimal cross-entropy risk for predicting $Y^*$ among all decision rules based on $(S,V)$. More generally, $\tau(V)$ is optimal for predicting $Y^*$ under a broad class of common loss functions (e.g., squared error).
Second, $\tau(V)$ satisfies calibration for predicting the desert decision: conditional on $\tau(V)$, the distribution of $Y^*$ does not depend on the sensitive attribute, 
so individuals with the same score $\tau(V)$ have the same probability of the favourable desert decision regardless of group membership \citep{chouldechova2017fair}. 
Third, the predicted decision based on $\tau(V)$ satisfies the fairness requirement by construction, because it does not depend on $S$. 
Taken together, using $\tau(V)$ allows us to pursue predictive accuracy for $Y^*$ while preserving fairness with respect to $S$, thereby avoiding the conventional utility--fairness trade-off that arises when predicting $Y$ under fairness constraints.

To quantify the degree of unfairness in the observed decision, we consider the probability that the observed decision differs from the desert decision, that is,
\[ \theta = \pr(Y\neq Y^*). \]
When $\theta>0$, the observed decision is subject to unfairness; larger values of $\theta$ indicate more severe unfairness. If $Y^*=1$ but $Y=0$, individuals are denied the favourable decision they deserve, which can be interpreted as discrimination; 
conversely, if $Y^*=0$ but $Y=1$, individuals receive more favourable outcomes than they deserve, suggesting preferential treatment. 
Thus, $\theta$ summarises the overall degree of unfairness in the observed decision. 
One may also consider more specific parameters, such as $\pr(Y=0\mid Y^*=1)$ to quantify discrimination and $\pr(Y=1\mid Y^*=0)$ to quantify preferential treatment.

The key distinction between our proposal and existing frameworks is the explicit introduction of the desert decision.
Our approach targets the prediction of $Y^*$ and uses $Y^*$ to characterise the mechanism of unfairness.
When unfairness is present, $Y^*$ and $Y$ are distinct variables. 
Conditional statistical parity is defined for the observed decision $Y$ or the predicted decision $\hat Y$ and can be assessed empirically, whereas Assumption~\ref{ass:fair} concerns the unobserved $Y^*$. Likewise, existing measures of unfairness typically rely on observed, predicted, or counterfactual decisions without reference to the desert decision, whereas our measures are grounded in $Y^*$ and directly evaluate discrepancies between the observed and desert decisions. For example, the demographic disparity $\pr(Y=1 \mid S=1) - \pr(Y=1 \mid S=0)$ captures differences in the observed decision across sensitive-attribute groups, but it does not reveal the extent to which the observed decision departs from what individuals rightfully deserve. Moreover, predicting $Y$ subject to imposed fairness constraints may yield an enforced balance rather than the decision that individuals rightfully deserve; 
the supplementary material provides an example comparing $\tau(V)$ with representative methods that target $Y$, illustrating that a single decision rule generally cannot deliver strong predictive performance for both $Y^*$ and $Y$ simultaneously.
Relative to existing frameworks, however, determining the decision rule $\tau(V)$ and the parameter $\theta$ is more challenging because both involve the unobserved variable $Y^*$. In the next section, we establish identification of $\tau(V)$ and $\theta$ by exploiting Assumption~\ref{ass:fair} together with additional assumptions on the unfairness mechanism of the observed decision $Y$.

\section{Identification} \label{sec:iden}

\subsection{Assumptions on the mechanism of unfairness}

We make the following assumptions about the   mechanism of unfairness in the observed decision. 
\begin{assumption}\label{ass:unfair}
(i) $\pr(Y=1\mid Y^*=0,S=0,V) =0$;
(ii) $\pr(Y=0\mid Y^*=1,S=1,V) =0$; and
(iii) $\pr(Y=0\mid Y^*=1,S=0,V) < 1$ and $\pr(Y=1\mid Y^*=0,S=1,V) < 1$.
\end{assumption}

\begin{assumption} \label{ass:aux}
The covariates vector $V$ includes  a binary auxiliary variable $Z$ such that  
(i) $Z \ind Y \mid (S,X,Y^*)$; and
(ii) $Z \nind Y^* \mid (S,X)$, where $X$ is the remainder of $V$, i.e.,  $V=(Z,X)$.
\end{assumption}

Here $\pr(Y=0\mid Y^*=1,S,V)$ and $\pr(Y=1\mid Y^*=0,S,V)$ encode the mechanism of unfairness: the former corresponds to discrimination and the latter corresponds to preferential treatment. Assumption~\ref{ass:unfair} (i)--(ii) rule out preferential treatment for the disadvantaged group and discrimination against the advantaged group; equivalently, the realised decision cannot be upgraded (from unfavourable $(Y^*=0)$ to favourable   $(Y=1)$)  for $S=0$
or downgraded (from unfavourable $(Y^*=1)$ to favourable   $(Y=0)$)  for $S=1$.
Although these restrictions may not hold exactly, they are often a reasonable approximation in many contexts: the primary sources of unfairness are typically discrimination against the disadvantaged group and preferential treatment for the advantaged group, captured by $\pr(Y=0\mid Y^*=1,S=0,V)$ and $\pr(Y=1\mid Y^*=0,S=1,V)$, respectively. 
These two quantities are unrestricted in our framework.
From a measurement error perspective, the unfairness mechanism can be viewed as misclassification probabilities linking the latent variable $Y^*$ to its mismeasured counterpart $Y$. Under this interpretation, Assumption~\ref{ass:unfair}(i)--(ii) corresponds to the one-sided misclassification condition, a common feature in the literature \citep{nguimkeu2019estimation,millimet2022accounting,mondal2024partial}. Condition (iii) is a standard positivity assumption that rules out extreme unfairness under which $Y=0$ for $S=0$ or $Y=1$ for $S=1$ almost surely.

Assumption~\ref{ass:aux} posits a binary auxiliary variable $Z$ satisfying the non-differentiality Condition (i), or exclusion restriction, 
namely that $Z$ does not affect the mechanism of unfairness.
Condition (ii) is a relevance requirement: $Z$ must be predictive of the desert decision $Y^*$ given $(S,X)$.
When the auxiliary variable $Z$ is multi-valued or continuous, it can provide additional restrictions that strengthen identification.
However, a binary $Z$ suffices for our identification analysis, and we focus on the binary case in this paper for simplicity.
In the measurement error literature, variables satisfying Assumption~\ref{ass:aux} are sometimes referred to as instruments or secondary measurements for the latent variable; see, for example, \citet{mahajan2006identification,lewbel2007estimation,hu2008identification,mondal2024partial}.

Under Assumptions~\ref{ass:unfair} and \ref{ass:aux},
the mechanism of unfairness is characterised by 
\begin{eqnarray}\label{eq:mecha}
\alpha(X) = \pr(Y=0\mid Y^*=1,S=0,X),\quad \beta(X) = \pr(Y=1\mid Y^*=0,S=1,X),
\end{eqnarray}
where $\alpha(X)$ quantifies the extent of discrimination against  the disadvantaged group ($S=0$) and $\beta(X)$ quantifies the degree of preferential treatment towards  the advantaged group ($S=1$),  conditional on covariates $X$.
The desert decision rule   can   be equivalently written as 
\[\tau(Z,X) = (1-Z)\tau_0(X) + Z\tau_1(X),\]
with $\tau_{z}(X) = \pr(Y^*=1\mid Z=z,X)$ for $z=0,1$.

\begin{figure}[h]
\centering
\subfloat[]{
\begin{tikzpicture}[scale=0.7,
->,
shorten >=2pt,
>=stealth,
node distance=1cm,
pil/.style={
->,
thick,
shorten =2pt,}
]
\node (S) at (1.5,-2) {$S$};
\node (Y2) at (3,0) {$Y^*$};
\node (Z) at (0,0) {$Z$};
\node (Y) at (4.5,-2) {$Y$};
\foreach \from/\to in {S/Y,Y2/Y,Z/Y2}
\draw (\from) -- (\to);
\end{tikzpicture}}
\quad
\subfloat[]{
\begin{tikzpicture}[scale=0.7,
->,
shorten >=2pt,
>=stealth,
node distance=1cm,
pil/.style={
->,
thick,
shorten =2pt,}
]
\node (S) at (1.5,-2) {$S$};
\node (Y2) at (3,0) {$Y^*$};
\node (Z) at (6,0) {$Z$};
\node (Y) at (4.5,-2) {$Y$};		
\foreach \from/\to in {S/Y,Y2/Y,Y2/Z}
\draw (\from) -- (\to);
\end{tikzpicture}}
\quad
\subfloat[]{
\begin{tikzpicture}[scale=0.7,
->,
shorten >=2pt,
>=stealth,
node distance=1cm,
pil/.style={
->,
thick,
shorten =2pt,}
]
\node (S) at (1.5,-2) {$S$};
\node (Y2) at (3,0) {$Y^*$};
\node (Z) at (0,0) {$Z$};
\node (Y) at (4.5,-2) {$Y$};
\foreach \from/\to in {S/Z,S/Y,Y2/Y,Z/Y2}
\draw (\from) -- (\to);
\end{tikzpicture}}
\quad
\subfloat[]{
\begin{tikzpicture}[scale=0.7,
->,
shorten >=2pt,
>=stealth,
node distance=1cm,
pil/.style={
->,
thick,
shorten =2pt,}
]
\node (S) at (1.5,-2) {$S$};
\node (Y2) at (3,0) {$Y^*$};
\node (Z) at (0,0) {$Z$};
\node (Y) at (4.5,-2) {$Y$};
\foreach \from/\to in {Z/S,S/Y,Y2/Y,Z/Y2}
\draw (\from) -- (\to);
\end{tikzpicture}}
\caption{Causal diagrams   satisfying Assumptions~\ref{ass:fair} and~\ref{ass:aux}, conditional on $X$.}\label{fig:dag}
\end{figure}
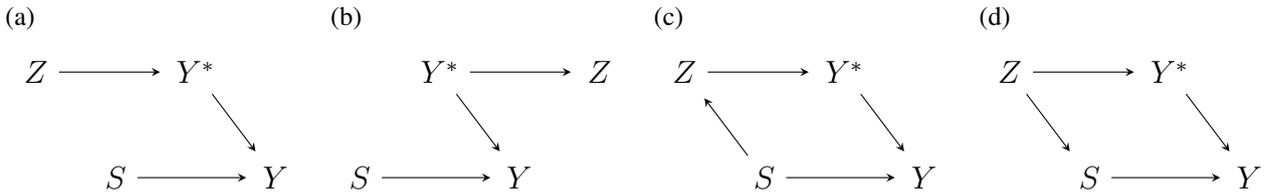

Figure~\ref{fig:dag} presents several causal diagrams corresponding to Assumptions~\ref{ass:fair} and~\ref{ass:aux}. 
In all diagrams, the sensitive attribute $S$ does not directly affect the desert decision $Y^*$, 
and the auxiliary variable $Z$ does not directly affect the observed decision $Y$. 
In panel~(a), $Z$ plays the role of an instrument that precedes $Y^*$ and is conditionally independent of $Y$ given $(X,Y^*)$. 
Panel~(b) depicts a setting in which $Z$ is a proxy of $Y^*$ subject to non-differential measurement error.
Panels~(c) and~(d) consider more general settings than panel~(a), allowing $S$ and $Z$ to be dependent.
The measurement error literature has also studied identification in models consistent with these causal diagrams; however, existing results either rely on parametric specifications \citep{nguimkeu2019estimation} or yield partial identification \citep{mondal2024partial}.
In contrast, we will establish nonparametric identification.
In the supplementary material, we further illustrate the proposed framework and assumptions using a generative model for the observed decision subject to unfairness.

We end this subsection with a practical example. We discuss the plausibility of Assumptions~\ref{ass:fair}--\ref{ass:aux} in this setting and further analyse the data in Section~\ref{sec:app}.

\begin{example}\label{ex:data}
In a study of racial discrimination in the labour market,
\citet{bertrand2004emily} sent fictitious resumes to help-wanted advertisements in newspapers.
The observed decision is whether an applicant receives a callback for an interview. The sensitive attribute is the randomly assigned Black- or White-sounding name on the resume, which serves as a proxy for the applicant's race. 
The resume content other than the name was sampled from templates constructed by the researchers and classified into two groups of high and low quality. 
For high quality resumes, additional features were selectively added to match job requirements, including extra experience, honours, or skills, while avoiding overqualification.
The covariates include city, occupation type, and job-seeker characteristics listed on the resume, such as labour market experience and educational attainment.
\end{example}

In this example, the object of interest is the unobserved desert decision that each applicant rightfully deserves based on the generated job-relevant information. Assumption~\ref{ass:fair} requires the desert decision to be independent of race given the other characteristics. Assumption~\ref{ass:unfair} is particularly compelling in this setting: \citet{bertrand2004emily} report no evidence of reverse discrimination and find  that callbacks are uniformly more responsive to resumes with White names than for  Black-sounding names, indicating discrimination against Black names and/or preferential treatment towards White names. 
To justify Assumption~\ref{ass:aux}, we use resume quality as the auxiliary variable. Resume quality is assigned and recorded by the researchers but is not explicitly revealed on the resume and therefore is not directly observed by employers. Consequently, conditional on the other resume covariates, resume quality should not affect the unfairness mechanism. At the same time, it is reasonable that resume quality is predictive of the desert decision.

\subsection{Nonparametric identification}

For notational convenience, we let $\mu_{sz}(X) = \pr(Y = 1\mid S=s,Z=z,X)$ for $(s,z)\in\{0,1\}^2$.
Assumptions~\ref{ass:fair}--\ref{ass:aux} suffice to identify the parameters of interest.
\begin{theorem}\label{thm:identification}
Under Assumptions~\ref{ass:fair}--\ref{ass:aux},
the joint distribution $\pr(S,Z,X,Y^*,Y)$ is identified and
\begin{align}
\tau_z(X) = T_z(X),\quad T_z(X) = \frac{ \mu_{0z}(X) \{\mu_{11}(X) - \mu_{10}(X)\}}{\mu_{01}(X)\{1- \mu_{10}(X)\} - \mu_{00}(X)\{1- \mu_{11}(X)\}}, \quad z=0,1. \label{eq:tau}
\end{align}
\end{theorem}

Theorem \ref{thm:identification} establishes   identification of the joint distribution and  provides a closed-form identification formula for the desert decision rule $\tau_z(X)$ under Assumptions~\ref{ass:fair}--\ref{ass:aux}, 
by noting that $T_z(X)$ is available from the observed data.
The identification formula $T_z(X)$ admits a bias-correction interpretation: it equals the observed probability $\mu_{0z}(X)$ multiplied by a correction factor that accounts for bias arising from unfairness.
In particular, $\mu_{0z}(X)$ is downward biased relative to $\tau_z(X)$ due to discrimination against the disadvantaged group. 
The correction factor can be viewed as a comparison of differences in $\mu_{sz}(X)$ across levels of $Z$, 
with the numerator given by the difference in $\mu_{1z}(X)$ and the denominator by the difference in $\mu_{0z}(X)\{1 - \mu_{1(1-z)}(X)\}$ across $z \in \{0,1\}$. 
Because the auxiliary variable $Z$ shifts the desert decision $Y^*$ without affecting the unfairness mechanism, this term isolates and corrects for the bias in $\mu_{0z}(X)$, thereby recovering the distribution of the desert decision.
The denominator of $T_z(X)$ is guaranteed to be nonzero under the relevance condition in Assumption~\ref{ass:aux}(ii).
An immediate consequence of Theorem~\ref{thm:identification} is that the degree of unfairness $\theta$ is identified.
The following result  enables identification and  estimation of $\theta$ using    $\tau(Z,X)$, $\alpha(X)$, and $\beta(X)$. 
\begin{theorem}\label{thm:para}
Under Assumptions~\ref{ass:fair}--\ref{ass:aux},
we can identify $\theta$ with
\begin{eqnarray}\label{eq:unfair}
\theta = E[(1-S)\tau(Z,X)\alpha(X) + S\{1-\tau(Z,X)\}\beta(X)].
\end{eqnarray}
\end{theorem}

The intuition   can be conveyed with a parameter-counting argument.
We start from the factorisation of the joint distribution  of $Y^*$ and $Y$ given $(S,Z,X)$:
\begin{align*}
\pr(Y^*,Y\mid S,Z,X) = \pr(Y^*\mid S,Z,X) \pr(Y\mid Y^*,S,Z,X),
\end{align*}
where $\pr(Y^*\mid S,Z,X)$ denotes the conditional distribution of the desert decision 
and $\pr(Y\mid Y^*,S,Z,X)$ characterises the mechanism of unfairness.
Without additional restrictions, for any fixed value $x$,  $\pr(Y^*\mid S,Z,X=x)$ involves four unknown parameters and $\pr(Y\mid Y^*,S,Z,X=x)$   eight unknown parameters.
By contrast, the observed data distribution yields only    four constraints,   encoded in
\begin{align}\label{eq:restriction}
\pr(Y=1\mid S,Z,X) =& \pr(Y=1\mid Y^*=0,S,Z,X)\pr(Y^*=0\mid S,Z,X) \nonumber\\
&+ \pr(Y=1\mid Y^*=1,S,Z,X)\pr(Y^*=1\mid S,Z,X),\quad (S,Z) \in \{0,1\}^2.
\end{align}
These four restrictions are insufficient for identification.
However, under Assumption~\ref{ass:fair},   $\pr(Y^*\mid S,Z,X)$  depends only on two parameters, $\tau_0(X)$ and $\tau_1(X)$;
under Assumption~\ref{ass:unfair},   $\pr(Y=1\mid Y^*=0,S=0,Z=z,X) =0$ and $\pr(Y=0\mid Y^*=1,S=1,Z=z,X) = 0$ for $z=0,1$,
and under Assumption~\ref{ass:aux}, $\pr(Y\mid Y^*,S,Z,X)$ further depends only on two parameters,  $\alpha(X)$ and $\beta(X)$ defined in Equation~\eqref{eq:mecha}.
Consequently,    \eqref{eq:restriction} can be rewritten as
\begin{eqnarray}\label{eq:mu}
\mu_{0z}(X)  =  \tau_z(X)\{1-\alpha(X)\},\quad \mu_{1z}(X)  =  \beta(X) + \tau_z(X)\{1-\beta(X)\},\quad z=0,1.
\end{eqnarray}
For any fixed value $x$, this system comprises four equations in four unknown quantities, which clarifies the role of the assumptions in  identification. 
Although the equations are nonlinear, the solution is unique in this setting; see the proof of Theorem~\ref{thm:identification} and the identification formula for $\alpha(X)$ and $\beta(X)$ in the supplementary material.

Theorem~\ref{thm:identification} does not rely on specific parametric models or impose functional form restrictions on $\tau_z(X)$,
and  the impact of covariates $X$ on   the desert decision and the unfairness mechanism can be arbitrary.
It opens the way to the estimation of the decision rule $\tau_z(X)$ and other parameters using various parametric or nonparametric models,
even though the desert decision $Y^*$ is unobserved.
Compared with existing methods that seek to predict the observed decision $Y$ under fairness constraints,
our decision rule $T_z(X)$ takes a markedly different form.
The difference arises from shifting the target of prediction from $Y$ to $Y^*$ and from invoking structural assumptions about the unfairness mechanism.
A related yet distinct line of work is the label debiasing method proposed by \citet{jiang2020identifying},
which similarly views the observed decision $Y$ as a biased proxy for an unobserved decision $Y^*$ (though without interpreting $Y^*$ as the desert decision)
and aims to recover the distribution of $Y^*$.
However, their framework relies on the assumption that the conditional distribution of $Y$ is the closest approximation to that of $Y^*$ among all candidates exhibiting the same level of bias,
whereas our assumptions on the unfairness mechanism can be more transparent and causally interpretable.
Moreover,
it is not entirely clear whether the distribution of $Y^*$ is identified under the assumption of \citet{jiang2020identifying},
whereas we provide rigorous identification results.

We   note that Assumptions~\ref{ass:fair}--\ref{ass:aux} impose restrictions on the observed data distribution. The following result states testable implications in terms of the observed functions $\mu_{sz}(X)$.

\begin{proposition}\label{prop:impl}
Under Assumptions~\ref{ass:fair}--\ref{ass:aux}, we have that
(i)~$\mu_{1z}(X) \geq \mu_{0z}(X)$ for $z\in\{0,1\}$;
(ii)~$\mu_{s1}(X) \neq \mu_{s0}(X)$ for $s\in\{0,1\}$;
and (iii)~$\mu_{11}(X) - \mu_{10}(X)$ and $\mu_{01}(X) - \mu_{00}(X)$ have the same sign.
\end{proposition}

If any of these restrictions are violated, then at least one of Assumptions~\ref{ass:fair}--\ref{ass:aux} must fail. However, even if the observed data are consistent with these restrictions, this does not establish that Assumptions~\ref{ass:fair}--\ref{ass:aux} hold. 
Their plausibility depends on domain-specific knowledge and must be assessed on a case-by-case basis. 
The logic is analogous to the testable implications of the assumptions underpinning instrumental variable methods for causal effect estimation \citep{angrist1996identification,balke1997bounds,kitagawa2015test}.
Sensitivity analysis   for Assumptions~\ref{ass:fair}--\ref{ass:aux} is therefore   important   for applying the proposed framework in practice,
which we will elaborate  in Section 5.

\section{Estimation} \label{sec:est}

\subsection{Estimation of the desert decision rule}\label{sec:tau}
In this section, we estimate the desert decision rule $\tau(Z,X)$, equivalently the functions $\tau_0(X)$ and $\tau_1(X)$. A natural approach is to first estimate the conditional probabilities $\mu_{sz}(X)$ for $(s,z)\in\{0,1\}^2$ and then plug in  these estimates into \eqref{eq:tau}. A key limitation, however, is that the resulting estimators of $\tau_0(X)$ and $\tau_1(X)$ are not guaranteed to lie in $[0,1]$ in finite samples, and hence may fail to produce valid probabilities.
We therefore adopt a nonparametric sieve maximum likelihood approach. 
For notational simplicity, we suppress the covariates $X$ when no confusion arises; for example, we write $\tau_0$, $\tau_1$, $\alpha$, $\beta$ for $\tau_0(X)$, $\tau_1(X)$, $\alpha(X)$, $\beta(X)$, respectively.
Let $\xi=(\tau_0,\tau_1,\alpha,\beta)$ and $\tilde\xi=(\tilde\tau_0,\tilde\tau_1,\tilde\alpha,\tilde\beta)$, where $\tilde\tau_0$, $\tilde\tau_1$, $\tilde\alpha$, and $\tilde\beta$ denote generic approximating functions. 
Let $O=(S,Z,X,Y)$ denote the observed data. 
The conditional log-likelihood of $Y$ given $(S,Z,X)$ is
\begin{eqnarray*}\label{eq:cri}
l(O;\tilde\xi) & = & Y\log \pr(Y=1\mid S,Z,X;\tilde\xi) + (1-Y)\log\{1-\pr(Y=1\mid S,Z,X;\tilde\xi)\},
\end{eqnarray*}
where
\begin{eqnarray}\label{eq:condprob}
\pr(Y=1\mid S,Z,X;\tilde\xi) &=& (1-S)(1-Z)\mu_{00}(X;\tilde\tau_0, \tilde\alpha) + (1-S)Z\mu_{01}(X;\tilde\tau_1, \tilde\alpha) \nonumber\\
&&+ S(1-Z)\mu_{10}(X;\tilde\tau_0, \tilde\beta) + SZ\mu_{11}(X;\tilde\tau_1, \tilde\beta),
\end{eqnarray}
with the component functions specified according to  \eqref{eq:mu},
i.e.,
\begin{eqnarray*}
\mu_{0z}(X;\tilde\tau_z, \tilde\alpha)  =  \tilde\tau_z(X)\{1-\tilde\alpha(X)\},\quad
\mu_{1z}(X;\tilde\tau_z, \tilde\beta)  =  \tilde\beta(X) + \tilde\tau_z(X)\{1-\tilde\beta(X)\},\quad z=0,1.
\end{eqnarray*}
The sieve maximum likelihood method estimates the parameters by maximising a criterion function; 
here, the population criterion function is $E\{l(O;\tilde\xi)\}$, which is uniquely maximised at the true parameter $\xi$.
Given $n$ independent and identically distributed copies of $O$.
The sieve maximum likelihood estimator $\hat\xi = (\hat\tau_0,\hat\tau_1,\hat\alpha,\hat\beta)$ is obtained by maximising the empirical criterion function over a sequence of approximating spaces $\mathcal F_n$,
i.e.,
\begin{align}\label{eq:problem}
\hat\xi = \arg\max_{\tilde\xi \in \mathcal F_n}\ \hat E\{l(O;\tilde\xi)\},
\end{align}
where $\hat E(\cdot)$ denotes the empirical mean operator.
Then, the estimator of   $\tau(Z,X)$ is
\[\hat\tau(Z,X) = (1-Z)\hat\tau_0(X) + Z\hat\tau_1(X).\]
For the sieve spaces $\mathcal F_n$, we employ a series logit specification, which is simple to implement. Let $\{\psi_j(X)\}_{j=1}^\infty$ be a sequence of basis functions (e.g., power series, splines, or wavelets), 
and define $\phi(X) = (\psi_1(X), \dots, \psi_{J_n}(X))^\t$ as the vector of the first $J_n$ basis functions,
where $J_n$ increases with the sample size $n$.
Denote $\expit(\cdot) = \exp(\cdot)/\{1+\exp(\cdot)\}$.
We approximate $\tau_0(X)$, $\tau_1(X)$, $\alpha(X)$, and $\beta(X)$ with logistic series expansions of the form  $\expit\{\gamma^\t \phi(X)\}$.

To establish the asymptotic properties of the estimators $\hat\xi$ and $\hat\tau$,
we introduce some notation and regularity conditions.
For a generic vector of functions $g(O) = (g_1(O),\dots,g_q(O))$,
let $\|g\|_2 = [\sum_{i=1}^q E\{g_i^2(O)\}]^{1/2}$ 
denote the $L_2$ norm with respect to the observed data distribution.
For a generic function $h(x)$,
let $\|h\|_{\infty,p}$ denote the Sobolev norm, with $p>0$ characterises the smoothness.
Let $\pi_{sz}(X) = \pr(S=s,Z=z\mid X)$ for $(s,z)\in\{0,1\}^2$ denote the   joint probabilities of the sensitive attribute and auxiliary variable given covariates.
We make the following conditions for the distribution and the sieve spaces,
which are standard in sieve estimation \citep{chen2007large}.

\begin{assumption}\label{ass:dist}
(i)~The vector of covariates $X \in \mathbb R^d$ has a compact support $\mathcal X = [0,1]^d$;
(ii)~Positivity:  
$\pi_{sz} \geq c$ for $(s,z)\in\{0,1\}^2$,
$c \leq \tau_0, \tau_1, \alpha, \beta \leq 1-c$,
and $|\tau_1 - \tau_0| \geq c$ for  some   $c>0$;
(iii)~Smoothness:
$\|\tau_0\|_{\infty, p}, \|\tau_1\|_{\infty, p}, \|\alpha\|_{\infty, p}, \|\beta\|_{\infty, p}< \infty$ for some $p>0$.
\end{assumption}

\begin{assumption}\label{ass:sieve}
(i)~The smallest eigenvalue of $E\{\phi(X)\phi(X)^\t\}$ is bounded away from zero for all $J_n$;
(ii)~letting $\mathcal H_n = \{\expit\{\gamma^\t \phi(X)\}, \gamma \in \mathbb R^{J_n}\}$,
the sieve spaces $\mathcal F_n$ are
\[\mathcal F_n = \{\tilde\xi = (\tilde\tau_0,\tilde\tau_1, \tilde\alpha, \tilde\beta): \tilde\tau_0,\tilde\tau_1, \tilde\alpha, \tilde\beta \in \mathcal H_n, c \leq \tilde\tau_0, \tilde\tau_1, \tilde\alpha, \tilde\beta \leq 1-c, \text{ and } |\tilde\tau_1-\tilde\tau_0|\geq c\};\]
(iii)~there exist an operator $\mathcal P_n$ such that $\mathcal P_n\xi \in \mathcal F_n$ and $\|\mathcal P_n\xi - \xi\|_2 = O(J_n^{-p/d})$.
\end{assumption}

\begin{theorem}\label{thm:est}
Under Assumptions~\ref{ass:fair}--\ref{ass:sieve},
letting $J_n = O(n^{d/(2p+d)})$,
we have
\begin{eqnarray*}
\|\hat\xi - \xi\|_2= O_p(n^{-p/(2p+d)}),\quad \|\hat\tau - \tau\|_2= O_p(n^{-p/(2p+d)}).
\end{eqnarray*}
\end{theorem}

The proof of Theorem~\ref{thm:est} adopts the local Rademacher complexity techniques in statistical learning theory \citep[see e.g.,][]{wainwright2019high},
which also provides finite sample error bounds.
In line with the classical bias--variance trade-off,
the estimation error can be decomposed into two parts:
an approximation error of order $J_n^{-p/d}$ and a standard deviation term of order $(J_n/n)^{1/2}$.
By choosing $J_n = O(n^{d/(2p+d)})$ to balance these two components,
the estimators achieve the best convergence rate $O_p(n^{-p/(2p+d)})$,
which matches the minimax optimal rate for standard nonparametric estimation \citep{tsybakov2009introduction}.
In addition to the sieve maximum likelihood estimator,
it is of great interest to employ other flexible or nonparametric estimation methods such as random forests, neural networks, and reproducing kernel Hilbert space methods.

We remark that, as in many fairness-enhancing methods, the proposed estimator $\hat\tau$ is obtained by solving an optimisation problem. 
A key difference is that the optimisation problem in \eqref{eq:problem} does not impose additional fairness constraints, whereas existing approaches typically encode fairness requirements via regularisations or explicit constraints and therefore trade utility against fairness. 
In our   framework, fairness is satisfied by construction when targeting the desert decision $Y^*$, thereby aligning predictive accuracy with fairness.

\subsection{Estimation of  the degree of unfairness}\label{sec:theta}
The parameter $\theta$ is a complicated functional of the observed data distribution, depending on   nuisance functions $\tau$, $\alpha$, and $\beta$ as shown in   \eqref{eq:unfair}.
After obtaining estimators $\hat\tau$, $\hat\alpha$, and $\hat\beta$, 
$\theta$ can be estimated by plugging in them into the sample analogue of \eqref{eq:unfair}. 
However,  the plug-in estimator need not be asymptotically normal, which complicates uncertainty quantification and  inference for $\theta$. 
We therefore develop an influence-function-based estimator  using semiparametric efficiency theory \citep{bickel1993efficient,tsiatis2006semiparametric} to obtain valid inference under standard regularity conditions.
We begin by deriving an influence function for $\theta$.

\begin{theorem}\label{thm:if}
Under Assumptions~\ref{ass:fair}--\ref{ass:aux},
an influence function for $\theta$ is
$\IF(O;\eta) = \varphi(O;\eta) -\theta$,
where $\eta = (\tau_0,\tau_1,\alpha,\beta,\pi_{00},\pi_{01},\pi_{10},\pi_{11})$ denotes the vector of nuisance functions,
and
\begin{eqnarray*}
\varphi(O;\eta) &=& (1-S)(1-Z)\{C_{00}(Y-\tau_0+\tau_0\alpha)+ \tau_0\alpha\}
+ (1-S)Z\{C_{01}(Y-\tau_1+\tau_1\alpha)+ \tau_1\alpha\} \\
&& +S(1-Z)\{C_{10}(Y-\beta-\tau_0+\tau_0\beta)+ \beta-\tau_0\beta\}
+SZ\{C_{11}(Y-\beta-\tau_1+\tau_1\beta)+ \beta-\tau_1\beta\},
\end{eqnarray*}
with $C_{sz}$ for $(s,z)\in\{0,1\}^2$ being functions of $X$ given in the supplementary material.
\end{theorem}

Theorem~\ref{thm:if} provides a closed-form influence function for $\theta$ under the identifying assumptions.
The influence function comprises   four components corresponding to four values of $(S,Z)$.
Relative  to   \eqref{eq:unfair},
each component contains an additional mean-zero augmentation term of the form,
$I(S=s,Z=z)C_{sz}(Y-\mu_{sz})$ for $(s,z)\in\{0,1\}^2$.
This construction is in the same spirt to the augmented inverse probability weighted estimator in causal inference and missing data literature \citep{bang2005doubly}.
Although the coefficients $C_{sz}$ appear to be algebraically  complicated,
they are functions obtained by arithmetic operations on the nuisance functions in $\eta$.
As a result, the first-order impact of nuisance estimation error is removed, and the remaining bias of the influence-function-based estimator is of second order under standard conditions.

Let $\hat\eta=(\hat\tau_0,\hat\tau_1,\hat\alpha,\hat\beta,\hat\pi_{00},\hat\pi_{01},\hat\pi_{10},\hat\pi_{11})$ denote the vector of nuisance estimators.
The proposed influence-function-based estimator is
\begin{eqnarray*}
\hat\theta = \hat E\{\varphi(O;\hat\eta)\}.
\end{eqnarray*}
The nuisance functions in $\eta$ can be partitioned into   two components:
$\xi = (\tau_0,\tau_1,\alpha,\beta)$ and the conditional probabilities $\pi_{sz}$ for $(s,z)\in\{0,1\}^2$.
In Section \ref{sec:tau},
we have described a  sieve maximum likelihood method for estimating $\xi$.
Estimation of $\pi_{sz}$   can be formulated as a multi-class classification problem with four classes indexed  by   $(s,z)\in \{0,1\}^2$.
We may apply standard classifiers such as multinomial logistic regression or more flexible methods. 
The following theorem establishes the asymptotic normality of the proposed estimator.

\begin{theorem}\label{thm:asynor}
Under the conditions of Theorem \ref{thm:est},
suppose that the nuisance estimators further satisfy that
(i) $ \hat\pi_{sz} \geq c$ for some $c>0$, $(s,z)\in \{0,1\}^2$;
(ii) $\| \hat\eta - \eta\|_2 = o_p(n^{-1/4})$;
and (iii) $\hat\eta$ and $\eta$ are in a Donsker class,
then we have
\begin{eqnarray*}
\hat\theta - \theta = \hat E\{\IF(O;\eta)\} + o_p(n^{-1/2}).
\end{eqnarray*}
\end{theorem}
Theorem~\ref{thm:asynor} implies that $\hat\theta$ is $n^{1/2}$-consistent and asymptotically normal, provided the nuisance estimators converge faster than $n^{-1/4}$. 
By the convergence rate in Theorem~\ref{thm:est}, 
this requirement for $(\hat\tau_0,\hat\tau_1,\hat\alpha,\hat\beta)$ holds when the smoothness parameter $p$ exceeds half the covariate dimension $d$. 
Rates faster than $n^{-1/4}$ are standard in influence-function-based estimation of complex functionals and can be achieved by a range of flexible methods, including random forests, neural networks, and ensemble procedures \citep{chernozhukov2018double}. Theorem~\ref{thm:asynor} also imposes a Donsker condition to control the complexity of the nuisance models \citep{van1996weak}. This condition can be relaxed, allowing more complex nuisance estimators, by using cross-fitting \citep[e.g.,][]{robins2008higher,chernozhukov2018double}.

Under the conditions of Theorem~\ref{thm:asynor}, the asymptotic variance of $\hat\theta$ is   $E\{\IF(O;\eta)^2\}$,
which can be consistently estimated with
$\hat\sigma^2 = \hat E[\{\varphi(O;\hat\eta) - \hat\theta\}^2].$
We can construct the   $1-\alpha$ confidence interval for $\theta$,
$
[\hat\theta - q_{\alpha/2}\hat\sigma n^{-1/2}, \hat\theta +  q_{\alpha/2}\hat\sigma n^{-1/2}],
$
where $q_{\alpha/2}$ denotes the $1-\alpha/2$ quantile of standard normal distribution function.

\section{Extensions: legitimate differential treatment and sensitivity analysis}\label{sec:sen}

\subsection{Incorporating legitimate differential treatment}
In this section, we conduct sensitivity analysis with respect to violations of the key identifying Assumptions~\ref{ass:fair}--\ref{ass:aux}. 
We begin with violations of the fairness Assumption~\ref{ass:fair}. 
In some settings, fairness considerations may justify providing stronger support to the disadvantaged group rather than applying a uniformly fair treatment rule, 
in order to  improve   circumstances and promote flourishing more rapidly.
For example, Rawls’ difference principle argues that social and institutional arrangements should work to the greatest benefit of the least advantaged \citep{rawls1971justice}. 
Practical instances include the WTO's ``special and differential treatment'' provisions designed to support developing countries \citep{Doha2001} and ``diversity, equity, and inclusion'' (DEI) policies intended to expand opportunities for historically disadvantaged groups \citep{kalev2006best}.

We therefore extend the proposed framework to more general settings in which the sensitive attribute $S$ may influence the desert decision $Y^*$ and may enter the decision rule. 
Let 
\[\tau_{sz}(X) = \pr(Y^*=1\mid S=s,Z=z,X),\quad  (s,z)\in \{0,1\}^2,\]
denote the desert decision rule that depends on the sensitive attribute.
The influence of $S$ on $Y^*$ conditional on $(Z,X)$ is quantified by $\kappa_z(X)=\tau_{1z}(X)-\tau_{0z}(X)$ for $z=0,1$. 
We assume that $\kappa_0$ and $\kappa_1$ are specified a priori by the decision-maker based on external knowledge, representing the level of legitimate support for the disadvantaged group. We have the following result.

\begin{theorem}\label{thm:v1}
Under Assumptions~\ref{ass:unfair} and \ref{ass:aux} and  given $(\kappa_0, \kappa_1)$,
if $\tau_{01}(1-\kappa_0) \neq \tau_{00}(1-\kappa_1)$,
then $\pr(S,Z,X,Y^*,Y)$ is identified and
\begin{align*}
\tau_{sz} = \frac{\mu_{0z}(\mu_{11} - \mu_{10}) +\kappa_0\mu_{0z}(1-\mu_{11}) - \kappa_1\mu_{0z}(1-\mu_{10}) }{\mu_{01}(1-\mu_{10}) - \mu_{00}(1-\mu_{11})} + s\kappa_z,\quad (s,z)\in\{0,1\}^2.
\end{align*}
\end{theorem}
Theorem~\ref{thm:v1} establishes identification of the joint distribution $\pr(S,Z,X,Y^*,Y)$ and provides an identification formula for $\tau_{sz}(X)$ in general settings that allow the sensitive attribute to have a specified influence on the desert decision. 
It also facilitates sensitivity analysis for assessing the robustness of the identification and estimation results in Sections~\ref{sec:iden}--\ref{sec:est} by specifying and varying the sensitivity parameters $\kappa_0$ and $\kappa_1$. 
Theorem~\ref{thm:identification} is a special case of Theorem~\ref{thm:v1} with $\kappa_0=\kappa_1=0$.
The condition $\tau_{01}(1-\kappa_0) \neq \tau_{00}(1-\kappa_1)$ is imposed to ensure that the denominator is nonzero.
Equivalently, it can be written as $\tau_{01}/(1-\tau_{11}) \neq \tau_{00}/(1-\tau_{10})$, representing a generalised relevance condition under which $\tau_{0z}/(1-\tau_{1z})$ varies with $z \in \{0,1\}$.
Under Assumptions~\ref{ass:unfair} and \ref{ass:aux}, a sufficient condition for this inequality is $\kappa_0 = \kappa_1$, that is, when the influence of $S$ on $Y^*$ is homogeneous across levels of $Z$.

The method in Section~\ref{sec:est} extends to estimation of $\tau_{sz}(X)$ by incorporating the prescribed level of legitimate differential treatment and modifying the criterion function accordingly. 
Given $\kappa_0$ and $\kappa_1$, the sieve maximum likelihood estimator $(\hat\tau_{00},\hat\tau_{01},\hat\alpha,\hat\beta)$ is obtained by solving the optimisation problem in \eqref{eq:problem}, with the component functions in \eqref{eq:condprob} replaced by
\begin{eqnarray*}
\mu_{0z}(X;\tilde\tau_{0z}, \tilde\alpha)  =  \tilde\tau_{0z}(X)\{1-\tilde\alpha(X)\},\quad
\mu_{1z}(X;\tilde\tau_{0z}, \tilde\beta)  =  \tilde\beta(X) +\{\tilde\tau_{0z}(X) + \kappa_z(X)\}\{1-\tilde\beta(X)\}.
\end{eqnarray*}
The resulting estimator of the decision rule is $\hat \tau_{sz}(X)=\hat\tau_{0z}(X) + s\kappa_z(X)$. 
To estimate the degree of unfairness $\theta$, 
we derive an influence function under the conditions of Theorem~\ref{thm:v1} and construct the corresponding estimator. Details are provided in the supplementary material.

\subsection{Sensitivity analysis against violations of  the unfairness mechanism}\label{sec:v23}

We next consider violations of Assumption~\ref{ass:unfair}. 
Let $\delta_0(X)=\pr(Y=1\mid Y^*=0,S=0,X)$ and $\delta_1(X)=\pr(Y=0\mid Y^*=1,S=1,X)$ denote, respectively, the probability of preferential treatment for the disadvantaged group and the probability of discrimination against the advantaged group. 
These quantities serve as sensitivity parameters that quantify departures from Assumption~\ref{ass:unfair}.

\begin{proposition}\label{thm:v2}
Under Assumptions~\ref{ass:fair} and \ref{ass:aux} and given $(\delta_0, \delta_1)$,
if $\alpha + \delta_0 <1$ and $\beta + \delta_1 <1$, 
then $f(S, Z, X, Y^*, Y )$ is identified  and 
\begin{align*}
\tau_z = \frac{(\mu_{0z}-\delta_0)(\mu_{11}-\mu_{10})}{\mu_{01}(1-\mu_{10}) - \mu_{00}(1-\mu_{11}) - \delta_0(\mu_{11}-\mu_{10}) - \delta_1(\mu_{01}-\mu_{00})},\quad z=0,1.
\end{align*}
\end{proposition}

The conditions $\alpha + \delta_0 < 1$ and $\beta + \delta_1 < 1$ ensure that the denominator is nonzero. Equivalently, they can be written as $\delta_0 < 1 - \alpha$ and $\beta < 1 - \delta_1$, implying that $Y$ and $Y^*$ are positively associated conditional on the other variables; 
in the present context, they mean that preferential treatment occurs with smaller probability than a fair favourable decision.
Analogous conditions are standard in the measurement error literature to ensure that an observed proxy is informative for the latent variable \citep{mahajan2006identification,lewbel2007estimation}.

Proposition~\ref{thm:v2} has two substantive implications. First, it establishes identification of the joint distribution and the desert decision rule $\tau_z(X)$ even when Assumption~\ref{ass:unfair} is violated, by incorporating the sensitivity parameters $\delta_0$ and $\delta_1$. 
It thus generalises Theorem~\ref{thm:identification}: under Assumption~\ref{ass:unfair}, $\delta_0=\delta_1=0$, and hence $\tau_z(X)=T_z(X)$. 
Although $\delta_0$ and $\delta_1$ are typically unknown in practice, Proposition~\ref{thm:v2} facilitates sensitivity analysis for violations of Assumption~\ref{ass:unfair} by specifying and varying these parameters.
Given $\delta_0$ and $\delta_1$, the sieve maximum likelihood estimator $\hat\xi=(\hat\tau_0,\hat\tau_1,\hat\alpha,\hat\beta)$ is obtained by solving the optimisation problem in \eqref{eq:problem}, with the component functions in \eqref{eq:condprob} replaced by
\begin{equation*}
\begin{aligned}
\mu_{0z}(X;\tilde\tau_z, \tilde\alpha)  = \delta_0(X)+ \tilde\tau_z(X)\{1-\delta_0(X)-\tilde\alpha(X)\},\\  
\mu_{1z}(X;\tilde\tau_z, \tilde\beta)  =  \tilde\beta(X) + \tilde\tau_z(X)\{1-\delta_1(X)-\tilde\beta(X)\}.
\end{aligned}
\end{equation*}
In the supplementary material, we derive an influence function for the degree of unfairness $\theta$ under the conditions of Proposition~\ref{thm:v2} and construct the corresponding estimator.

Second, Proposition~\ref{thm:v2} characterises the robustness of    $T_z(X)$  in Theorem~\ref{thm:identification} that is constructed under Assumption~\ref{ass:unfair} to violations of that assumption. 
In the supplementary material, we quantify the bias of $T_z(X)$ in terms of the sensitivity parameters when Assumption~\ref{ass:unfair} is violated. 
In particular, we show that when $\delta_0$ and $\delta_1$ are small,
\begin{align*}
T_z - \tau_z \approx
\frac{\delta_0(1-\tau_z)}{1-\alpha} -  \frac{\delta_1\tau_z}{1-\beta}.
\end{align*}
Consequently, the bias of $T_z(X)$ is approximately linear in these sensitivity parameters, suggesting that $T_z(X)$ is robust to mild or moderate departures from Assumption~\ref{ass:unfair}.

Regarding violations of Assumption~\ref{ass:aux}, 
let $\alpha_z(X)=\pr(Y=0\mid Y^*=1,S=0,Z=z,X)$ and $\beta_z(X)=\pr(Y=1\mid Y^*=0,S=1,Z=z,X)$ for $z=0,1$ denote the unfairness mechanism. 
These functions generalise $\alpha(X)$ and $\beta(X)$ in Section~\ref{sec:iden} by allowing the mechanism to vary with $Z$, thereby capturing violations of the non-differentiality requirement. 
Let $\zeta_0(X) = \{1-\alpha_1(X)\}/\{1-\alpha_0(X)\}-1$ and $\zeta_1(X) = \{1-\beta_1(X)\}/\{1-\beta_0(X)\}-1$ denote the
the differential impact of $Z$ on the unfairness mechanism and serve as sensitivity parameters encoding departures from Assumption~\ref{ass:aux}.

\begin{proposition}\label{thm:v3}
Under Assumptions~\ref{ass:fair} and \ref{ass:unfair} and given $(\zeta_0, \zeta_1)$,
if $\tau_1\neq \tau_0$ and $0 \leq \alpha_z, \beta_z < 1$ for $z=0,1$,
then $\pr(S,Z,X,Y^*,Y)$ is identified and
\begin{align*}
\tau_{z} = \frac{(1+\zeta_0)^{1-z}\mu_{0z}\{\mu_{11}-\mu_{10}+\zeta_1(1-\mu_{10})\} }{(1+\zeta_1)\mu_{01}(1-\mu_{10}) - (1+\zeta_0)\mu_{00}(1-\mu_{11})},\quad z=0,1.
\end{align*}
\end{proposition}

Proposition~\ref{thm:v3} establishes identification of the joint distribution $\pr(S,Z,X,Y^*,Y)$ and the desert decision rule $\tau_z(X)$ under violations of Assumption~\ref{ass:aux} by incorporating the sensitivity parameters $\zeta_0$ and $\zeta_1$. 
It generalises Theorem~\ref{thm:identification}: under Assumption~\ref{ass:aux}, $\zeta_0=\zeta_1=0$, and hence $\tau_z(X)=T_z(X)$. Sensitivity analysis can therefore be conducted by specifying and varying $\zeta_0$ and $\zeta_1$.
Given $\zeta_0$ and $\zeta_1$, 
we obtain the sieve maximum likelihood estimator $\hat\xi=(\hat\tau_0,\hat\tau_1,\hat\alpha_0,\hat\beta_0)$ by solving the optimisation problem in \eqref{eq:problem}, with the component functions in \eqref{eq:condprob} replaced by
\begin{equation*}
\begin{gathered}
\mu_{0z}(X;\tilde\tau_z,\tilde\alpha_0)
= \{1+\zeta_0(X)\}^z \tilde\tau_z(X)\{1-\tilde\alpha_0(X)\},\\
\mu_{1z}(X;\tilde\tau_z,\tilde\beta_0)
= 1- \{1+\zeta_1(X)\}^z \{1-\tilde\tau_z(X)\}\{1-\tilde\beta_0(X)\}.
\end{gathered}
\end{equation*}
In the supplementary material, 
we derive an influence function for $\theta$ under the conditions of Proposition~\ref{thm:v3} and construct the corresponding estimator. 
We also show that, when $\zeta_0$ and $\zeta_1$ are small, the bias of $T_z(X)$ is approximately linear in these parameters, suggesting robustness to mild violations of Assumption~\ref{ass:aux}.
Moreover, when both Assumptions~\ref{ass:unfair} and~\ref{ass:aux} are violated, the bias of $T_z(X)$ is also quantified and shown to be approximately linear in the magnitude of the violations.
The robustness is further supported by the simulation results in Section~\ref{sec:sim}.

\section{Simulation} \label{sec:sim}

We evaluate the performance of the proposed methods via numerical simulations. 
We generate two covariates $X=(X_1,X_2)$, a binary sensitive attribute $S$, a binary auxiliary variable $Z$, a binary desert decision $Y^*$, and a binary observed decision $Y$ from the following model:
\begin{equation*}
\begin{gathered}
X_1\ind X_2 \text{ and } X_1, X_2 \thicksim {\rm U}(0,1),\\
S\ind Z\mid X, \quad \pr(S=1\mid X) = \expit(2-2\sin X_1-2X_2), \quad \pr(Z=1\mid X) = \expit(1-X_1-\sin X_2),\\
\pr(Y^*=1\mid S,Z,X) = (1-Z)\expit(-3+5X_1+\sin X_2) + Z\expit(-3+X_1+6\sin X_2),\\
\pr(Y=0\mid Y^*=1,S,Z,X) = (1-S)\expit\{-1 - \sin X_1 + 2 \exp(-X_2)\} + S\delta,\\
\pr(Y=1\mid Y^*=0,S,Z,X) = (1-S)\delta + S\expit\{-1+2 \exp(-X_1)-X_2\},
\end{gathered}
\end{equation*}
The above setting satisfies Assumptions~\ref{ass:fair} and \ref{ass:aux},
with $\tau_0(X) =  \expit(-3+5X_1+\sin X_2)$,
$\tau_1(X)= \expit(-3+X_1+6\sin X_2)$,
$\alpha(X) = \expit\{-1 - \sin X_1 + 2 \exp(-X_2)\}$,
$\beta(X) = \expit\{-1+2 \exp(-X_1)-X_2\}$.
It further implies $\pr(Y=0\mid Y^*=1,S=1,Z,X) = \pr(Y=1\mid Y^*=0,S=0,Z,X) = \delta$.
The sensitivity parameter $\delta$ quantifies departures from  Assumption~\ref{ass:unfair}:
it holds for $\delta=0$ but is violated for $\delta>0$,
with larger values indicating more severe violations. 
We consider three values,   $\delta\in \{0, 0.05, 0.1\}$.

We apply the sieve approach proposed in Section~\ref{sec:tau} to obtain the estimator $\hat\tau$. 
To estimate $\xi=(\tau_0,\tau_1,\alpha,\beta)$, we solve the optimisation problem in \eqref{eq:problem} using polynomial basis functions of order $3$. 
The conditional probabilities $\pi_{sz}$ for $(s,z)\in\{0,1\}^2$ are estimated using a standard series logit specification with the same polynomial basis. 
We refer to the proposed method as DSD, as decisions are made based on $\hat\tau$, which targets the desert decision $Y^*$.
We estimate the degree of unfairness $\theta$ and construct confidence intervals using the influence-function-based method proposed in Section~\ref{sec:theta}.
For comparison, we implement the following four methods:
\begin{itemize}
\item[(i)] UML---an unconstrained  machine learning predictor of the observed decision $Y$ based on $(S,Z,X)$, without fairness adjustment;
\item[(ii)] FTU---the fairness-through-unawareness method that excludes $S$ and predicts $Y$ from the nonsensitive variables $(Z,X)$;
\item[(iii)] MLC---a machine learning predictor of $Y$ based on $(S,Z,X)$ subject to a fairness constraint enforcing a zero causal effect of $S$ on $Y$ \citep{nabi2024statistical};
\item[(iv)] LD---the label debiasing method based on reweighting \citep{jiang2020identifying}.
\end{itemize}
Implementation details for competing methods are provided in the supplementary material.

\begin{figure}[H]
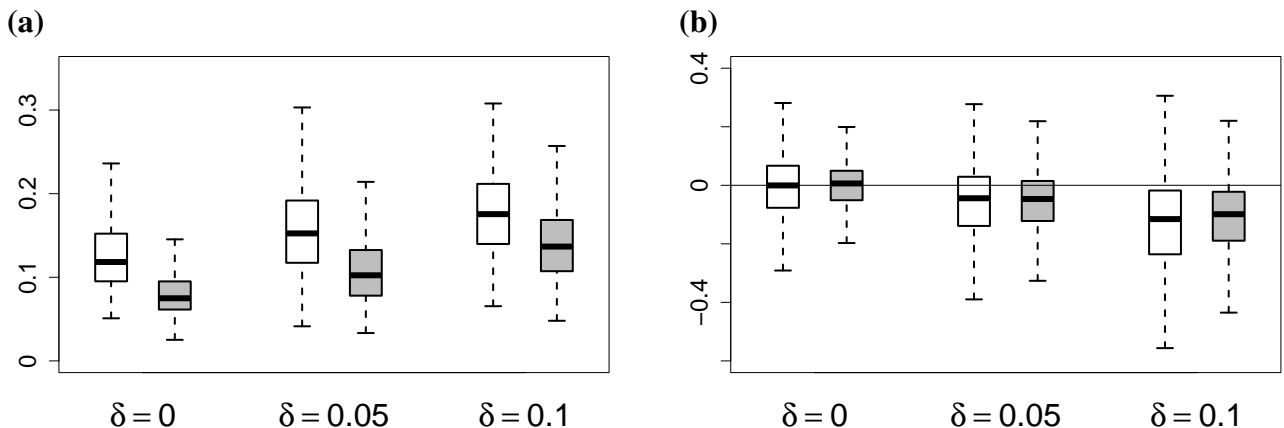

\graphicspath{{./Rscripts/Simulation/Results/delta/}}
\centering
\begin{subfigure}{0.48\textwidth}
\centering
\begin{overpic}[scale=0.5]{error_tau.pdf}
\put(1,64){\textbf{(a)}}
\end{overpic}
\end{subfigure}
\hfill
\begin{subfigure}{0.48\textwidth}
\centering
\begin{overpic}[scale=0.5]{bias_theta.pdf}
\put(1,64){\textbf{(b)}}
\end{overpic}
\end{subfigure}
\caption{(a) Estimation error of $\hat\tau$ and (b) bias of $\hat\theta$.}\label{fig:bias}
\begin{justify}
Note: White boxes are for   sample size $n=2000$ and grey for $4000$.
\end{justify}
\end{figure}

For each setting, we replicate $1000$ simulations at sample sizes $n = 2000, 4000$.
Figure~\ref{fig:bias}(a) reports the estimation error of the proposed estimator $\hat\tau$, measured by the $L_2$ norm $\|\hat\tau-\tau\|_2$ and evaluated on an independent test set of size $10^6$ generated from the same model. 
When $\delta=0$, so that Assumptions~\ref{ass:fair}--\ref{ass:aux} hold, the estimator has small error that decreases as the sample size increases. 
When $\delta>0$, the error increases with $\delta$ due to violations of Assumption~\ref{ass:unfair}. 
The estimator remains relatively stable for moderate values of $\delta$, but the error can become substantial under more severe violations of Assumption~\ref{ass:unfair}. 
The estimation errors for $\hat\tau_0$, $\hat\tau_1$, $\hat\alpha$, and $\hat\beta$ exhibit similar patterns, 
and we report these results in the supplementary material.

Table~\ref{tbl:auc} reports the area under the receiver operating characteristic curve (AUC) for each method when predicting $Y^*$ and $Y$. 
For predicting the desert decision $Y^*$, 
the proposed DSD method attains the highest AUC across all combinations of $n$ and $\delta$, 
whereas the unconstrained machine learning predictor UML performs worst. 
In contrast, for predicting the observed decision $Y$, UML achieves the highest AUC, while the DSD method performs worst. 
The fairness-oriented methods FTU, MLC, and LD reduce AUC for $Y$ relative to UML, consistent with an utility--fairness trade-off, 
and they also reduce AUC for $Y^*$ relative to the DSD method. 
Taken together, these results indicate a trade-off between predicting $Y$ and predicting $Y^*$ when the observed decision is affected by unfairness, 
so that a single method generally cannot achieve optimal performance for both targets.
Notably, the supplementary material presents an example in which the AUC values of UML, FTU, and MLC for predicting $Y^*$ fall below $0.5$, 
indicating performance worse than random guessing. 
When the identifying assumptions are plausible, 
these findings support using $\hat\tau$ to guide decision-making rather than methods that target the observed decision.

Figure~\ref{fig:bias}(b) reports the bias of the proposed estimator $\hat\theta$, and Table~\ref{tbl:cvp} reports the coverage probability of the corresponding confidence interval across settings with varying $\delta$. 
When $\delta=0$, so that Assumptions~\ref{ass:fair}--\ref{ass:aux} hold, the bias is small and decreases with sample size, and the nominal $95\%$ confidence interval attains coverage close to $0.95$. 
When $\delta>0$, Assumption~\ref{ass:unfair} is violated; the bias increases with $\delta$, and coverage falls substantially below $0.95$ for larger values (e.g., $\delta=0.1$). 
These results indicate that the proposed estimation and inference procedure for $\theta$ is valid under the identifying assumptions and is reasonably robust to mild violations, but may be unreliable under more severe departures.

We also conduct additional simulations to assess sensitivity to violations of Assumptions~\ref{ass:fair} and \ref{ass:aux}, with results reported in the supplementary material due to space constraints. 
The proposed estimators exhibit small error or bias under mild or moderate violations, but performance can deteriorate under more severe departures. 
In practice, we therefore recommend sensitivity analysis to evaluate robustness to potential violations of the identifying assumptions.

\begin{table}[H]
\centering
\caption{AUC values for different methods in predicting $Y^*$ and $Y$}\label{tbl:auc}
\begin{tabular}{llccccccccccc}
\hline
& & \multicolumn{5}{c}{$Y^*$} & &  \multicolumn{5}{c}{$Y$} \\
\multicolumn{2}{c}{Prediction methods:}  & DSD & UML & FTU & MLC & LD & & DSD & MLC & FTU & MLC & LD  \\
\hline
\multirow{2}*{$\delta=0$}  
& $n= 2000$ &  0.803 & 0.604 & 0.757 & 0.724 & 0.789 && 0.603 & 0.768 & 0.624 & 0.622 & 0.671 \\
& $n= 4000$ & \textbf{0.820} & 0.603 & 0.766 & 0.735 & 0.795  && 0.606 & 0.772 & 0.628 & 0.628 & 0.676 \\
\multirow{2}*{$\delta=0.05$}
& $n= 2000$ & 0.787 & 0.608 & 0.751 & 0.716 & 0.779 && 0.593 & 0.738 & 0.614 & 0.614 & 0.656 \\
& $n= 4000$ & \textbf{0.812} & 0.607 & 0.761 & 0.729 & 0.789 && 0.597 & 0.743 & 0.618 & 0.621 & 0.662 \\
\multirow{2}*{$\delta=0.1$}
& $n= 2000$ & 0.776 & 0.610 & 0.744 & 0.706 & 0.767 && 0.582 & 0.709 & 0.603 & 0.606 & 0.641 \\
& $n= 4000$ & \textbf{0.803} & 0.612 & 0.756 & 0.721 & 0.779 && 0.587 & 0.715 & 0.608&  0.613 & 0.647 \\
\hline
\end{tabular}
\end{table}

\begin{table}[H]
\centering
\caption{Coverage rate of the $95\%$ confidence interval for $\theta$}\label{tbl:cvp}
\begin{tabular}{lcccccc}
\hline
& & \multicolumn{3}{c}{$\delta$}\\
& & $0$ & $0.05$  & $0.1$ \\
\hline
$n=2000$ & & 0.954 & 0.938 & 0.851\\
$n=4000$ & & 0.961 & 0.914 &0.836\\
\hline
\end{tabular}
\end{table}

\section{Application} \label{sec:app}
We apply the proposed methods to data from the field experiment of \citet{bertrand2004emily}, introduced in Example~\ref{ex:data}. The data are publicly available from \url{https://doi.org/10.3886/E116023V1}. The sample size is $n=4{,}870$. 
The study evaluates the impact of an applicant’s race on labour market outcomes. 
Specifically, fictitious resumes with randomly assigned Black- or White-sounding names were sent to job advertisements in newspapers, and callback for an interview was recorded.
A direct examination of the data shows that the callback rate is $0.0965$ for resumes with white-sounding names and $0.0645$ for those with Black-sounding names. 
The estimated difference is $0.0320$, with a $95\%$ confidence interval of $(0.0168,0.0473)$.
\citet{bertrand2004emily} highlighted that White names receive nearly $50\%$ more callbacks.
These results indicate significantly fewer callbacks for resumes with Black-sounding names, consistent with the presence of unfairness in this setting.

By re-analysing the data using our methods, we aim to recover a fair decision rule that targets the desert decision, quantify the degree of unfairness in the observed decision, and examine the underlying unfairness mechanism.
Let $S$ and $Z$ denote the randomly assigned White- or Black-sounding name and resume quality, with $S=1$ for a White-sounding name and $S=0$ for a Black-sounding name, and $Z=1$ for high quality and $Z=0$ for low quality. 
Let $Y=1$ indicate a callback and $Y=0$ indicate no callback. 
We use resume quality as the auxiliary variable. 
Covariates $X$ include the number of prior jobs, years of work experience, and ten discrete variables: educational attainment, honours received, volunteer experience, military experience, presence of gaps in the resume, work during schooling, computer skills, special skills, city of the job, and occupation type.
For this application, the plausibility of Assumptions~\ref{ass:fair}--\ref{ass:aux} was discussed and justified in Example~\ref{ex:data}. 

The parameters of interest include the target desert decision rule $\tau(Z,X)=\pr(Y^*=1\mid Z,X)$, the degree of unfairness $\theta=\pr(Y\neq Y^*)$, and the unfairness mechanism $\alpha(X) = \pr(Y=0\mid Y^*=1,S=0,X)$, $\beta(X) = \pr(Y=1\mid Y^*=0,S=1,X)$,
We apply the methods in Section~\ref{sec:est} to estimate these quantities. 
Specifically, $\tau_0$, $\tau_1$, $\alpha$, and $\beta$ are estimated via the sieve maximum likelihood approach using polynomial basis functions of order $3$, and $\pi_{sz}$ for $(s,z)\in \{0,1\}^2$ are estimated using a standard series logit specification with the same polynomial basis.
To obtain a classifier based on $\hat\tau$ that preserves the overall rate of favourable decisions (i.e., the total number of callbacks), we define
\[\hat Y^* = I\{\hat\tau(Z,X) >  t^*\} \text{ with }
 t^* = \sup\{t \in [0,1]: \hat E[I\{\hat\tau(Z,X) \geq t\}] \geq \hat E(Y) \},\]
where $\hat E(Y)=0.0805$. 
For comparison, we implement the four methods UML, FTU, MLC, and LD introduced in Section~\ref{sec:sim}. 
For each competing method, we construct an analogous classifier that preserves the overall rate of favourable decisions, 
yielding corresponding predictions. 
Implementation details are provided in the supplementary material.

Tables~\ref{tbl:pred} and~\ref{tbl:pred-black} report the conditional proportions of decisions predicted by the UML, FTU, MLC, and LD methods given $\hat Y^*$ among all resumes and resumes with Black-sounding names, respectively, thereby comparing the proposed predictions with those from the four competing methods. 
When the proposed DSD method predicts no callback $(\hat Y^*=0)$, 
the competing methods agree with this prediction with probabilities close to one. 
In contrast, when the DSD method predicts a callback $(\hat Y^*=1)$, the competing methods often disagree. 
Among them, UML exhibits the largest discrepancy from the DSD method, whereas the remaining three methods show more moderate differences.
For resumes with Black-sounding names, a large share of applicants predicted a callback by the DSD method $(\hat Y^*=1)$ are predicted by the UML, MLC and LD methods to receive no callback, with proportions even exceeding 0.5. 
This discrepancy indicates that the competing methods may fail to recognise a considerable proportion of deserving individuals from the disadvantaged group compared with the DSD method.
In the supplementary material, we also compare the DSD method with competing methods when the resume quality $Z$ is not used for prediction; the results are similar.

\begin{figure}[H]
\graphicspath{{./Rscripts/Lakisha/Results/}}
\centering
\includegraphics[scale=0.55]{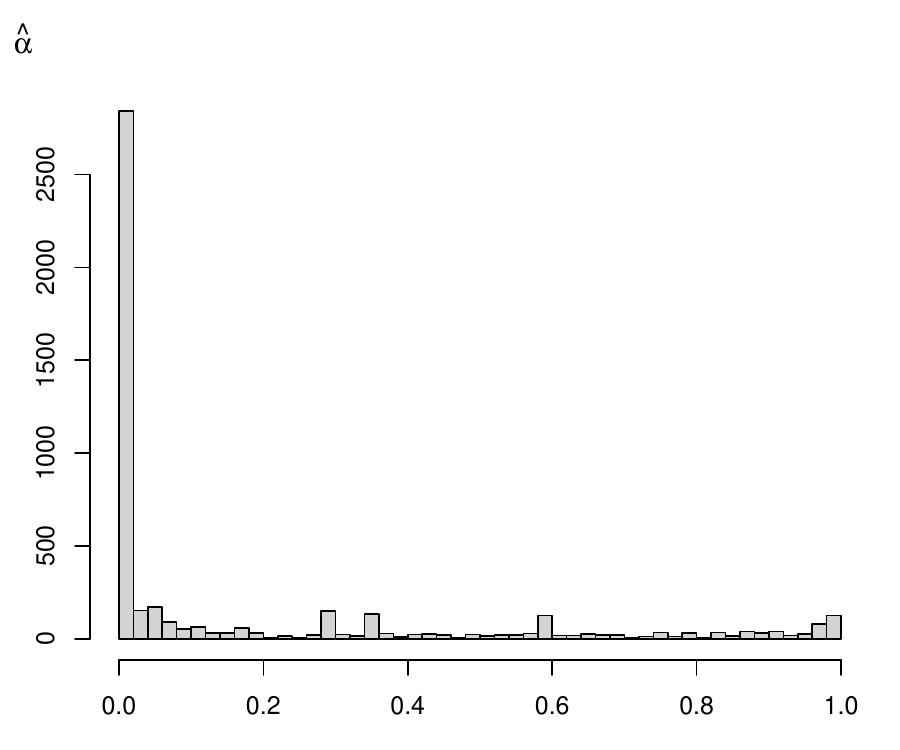}
\includegraphics[scale=0.55]{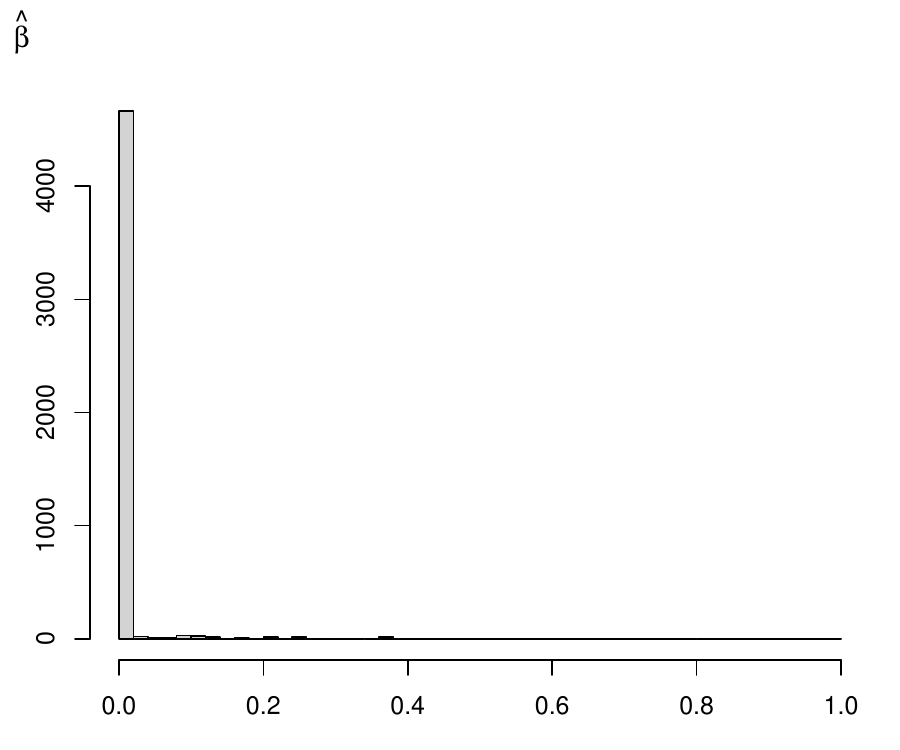}
\caption{Histograms of the estimated unfairness mechanism $\hat\alpha$ and $\hat\beta$.}
\label{fig:hist}
\end{figure}

\begin{figure}[H]
\graphicspath{{./Rscripts/Lakisha/Results/}}
\centering
\includegraphics[scale=0.55]{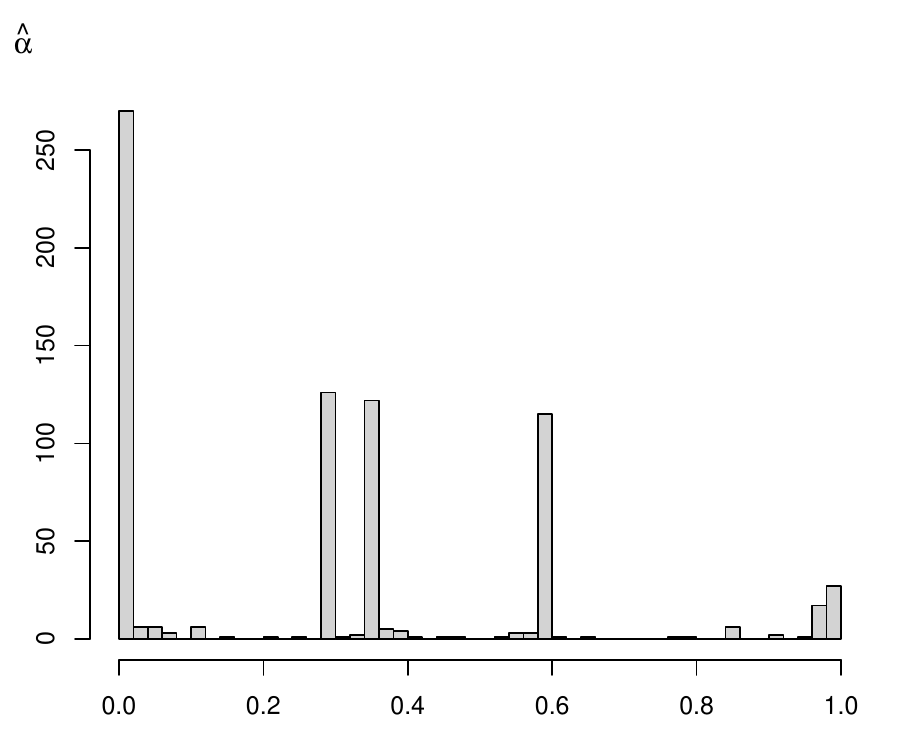}
\caption{Histogram of $\hat\alpha$ for applicants targeting sales representative positions.}
\label{fig:salesrep}
\end{figure}

The estimated degree of unfairness is $\hat\theta=0.0161$, 
which is significantly nonzero with a $95\%$ confidence interval of $(0.0085,\ 0.0236)$. 
Figure~\ref{fig:hist} displays histograms of the estimated values $\hat\alpha(X_i)$ and $\hat\beta(X_i)$ across applicants. 
For $\hat\beta(X_i)$, almost all values  concentrate around zero, 
but for $\hat\alpha(X_i)$, in addition to the concentration around zero, 
there exist a number of units with nonzero or large values.
Specifically, a distinct cluster appears around 0.6 for approximately 100--200 individuals. 
Further examination reveals that this cluster consists almost entirely of applicants targeting sales representative positions 
and constitutes nearly one‑third of nonzero values in this subgroup, as shown in Figure~\ref{fig:salesrep}. 
This pattern is consistent with existing discrimination theories \citep{darity1998evidence,becker2010economics} concerning the nature of such occupations, which typically involve extensive customer interaction and may therefore leave greater scope for discriminatory behaviour.
This finding suggests that the degree of unfairness is heterogeneous, and certain occupations may be associated with higher level of discrimination, 
highlighting the importance of moving beyond aggregate measures to understand how the nature of unfairness varies across subgroups.
Taken together, these results indicate statistically significant unfairness in the observed decisions, 
with the dominant contribution arising from discrimination against Black applicants, particularly for certain occupations. 
Preferential treatment towards White applicants appears comparatively mild.

We further conduct sensitivity analyses to assess robustness to potential violations of Assumptions~\ref{ass:fair}--\ref{ass:aux}; details are provided in the supplementary material. 
The results indicate that the predicted decision $\hat Y^*$ and the estimated degree of unfairness $\hat\theta$ are relatively insensitive to mild or moderate departures from these assumptions, whereas more severe violations can lead to substantial changes.

\begin{table}[H]
\centering
\caption{Conditional proportions of predicted decisions among all resumes}\label{tbl:pred}
\begin{tabular}{cccccccccccc}
\hline
& \multicolumn{2}{c}{UML} & & \multicolumn{2}{c}{FTU} & & \multicolumn{2}{c}{MLC} & & \multicolumn{2}{c}{LD}\\
DSD & 0 & 1 & & 0 & 1 & & 0 & 1 & & 0 & 1 \\
\hline
0 & $ 0.9665$ & $0.0335$ & & $0.9736$ & $0.0264$ & & $0.9692$ & $0.0308$ & & $ 0.9692$ & $0.0308$ \\
1 & $\textbf{0.3852}$ & $\textbf{0.6148}$ & & $ 0.3010$ & $0.6990$ & & $0.3520$ & $0.6480$ & & $0.3520$ & $0.6480$ \\
\hline
\end{tabular}
\end{table}

\begin{table}[H]
\centering
\caption{Conditional proportions of predicted decisions among
resumes with Black-sounding names}\label{tbl:pred-black}
\begin{tabular}{cccccccccccc}
\hline
& \multicolumn{2}{c}{UML} & & \multicolumn{2}{c}{FTU} & & \multicolumn{2}{c}{MLC} & & \multicolumn{2}{c}{LD}\\
DSD & 0 & 1 & & 0 & 1 & & 0 & 1 & & 0 & 1 \\
\hline
0 & $0.9933$ & $0.0067$ & & $0.9767$ & $0.0233$ & & $0.9781$ & $0.0219$ & & $0.9790$ & $0.0210$ \\
1 & $\textbf{0.6535}$ & $\textbf{0.3465}$ & & $0.3366$ & $0.6634$ & & $0.5149$ & $0.4851$ & & $0.5149$ & $0.4851$ \\
\hline
\end{tabular}
\end{table}

\setstretch{1.9}
\section{Discussion} \label{sec:dis}

We establish a framework for characterising and addressing fairness that builds on the notion of the desert decision. Although some readers may view this construct as metaphysical or unverifiable, we argue that introducing the desert decision can deepen our understanding of fairness and motivate the development of statistical methods that are useful in practice, including the approach proposed here.

Building on the desert decision, we define parameters for assessing and achieving fairness, establish identification under transparent and causally interpretable assumptions, 
and develop corresponding estimation  procedures. 
In contrast to most existing approaches that target the observed decision while imposing fairness constraints, 
our framework targets the desert decision directly. 
To achieve identification of the desert decision rule, we rely on assumptions on the fairness property of the desert decision and the unfairness mechanism generating the observed decision, 
which differs fundamentally from approaches that define target decision rules through constrained statistical learning. 
For estimation, we propose a sieve maximum likelihood estimator for the desert decision rule $\tau$ and the unfairness mechanism $(\alpha,\beta)$, and an influence-function-based estimator of the degree of unfairness $\theta$. 
We further develop sensitivity analysis procedures that allow for departures from Assumptions~\ref{ass:fair}--\ref{ass:aux}; 
it remains of interest to relax these assumptions further. 
Several extensions merit investigation in the future, including flexible estimation of $\tau$ in high-dimensional settings and more precise inference for $\theta$ via the efficient influence function and the corresponding semiparametrically efficient estimator.

Our proposal also contributes to the measurement error literature. 
Under this interpretation, the desert decision is an unobserved “true” variable, while the observed decision is a potentially misclassified proxy. 
To the best of our knowledge, our identification results under Assumptions~\ref{ass:fair}--\ref{ass:aux} provide new insights for binary choice models with misclassification. 
It would also be of interest to develop partial identification results when these assumptions are relaxed (see, e.g., \citealp{molinari2008partial, mondal2024partial}). 
As ongoing work, we are extending our identification arguments beyond binary variables to accommodate more general outcomes and decision spaces.

\section*{Supplementary material}
Supplementary material online includes
a generative model for the observed decision subject to unfairness,
characterisations of the bias of $T_z(X)$ under violations of the unfairness mechanism,
sensitivity analysis procedures for the degree of unfairness $\theta$,
proof of theorems and important results,
implementation details of comparing methods,
and additional results for the simulation and application studies.

\setstretch{1.7}

\bibliographystyle{chicago}
\bibliography{CausalMissing}

\end{document}